\documentclass[fleqn,usenatbib]{mnras}

\usepackage{newtxtext,newtxmath}
\usepackage[T1]{fontenc}

\DeclareRobustCommand{\VAN}[3]{#2}
\let\VANthebibliography\thebibliography
\def\thebibliography{\DeclareRobustCommand{\VAN}[3]{##3}\VANthebibliography}

\usepackage{graphicx}	% Including figure files
\usepackage{amsmath}	% Advanced maths commands
\usepackage{newtxtext,newtxmath}

\usepackage{multicol}        % Multi-column entries in tables
\usepackage{bm}		% Bold maths symbols, including upright Greek
\usepackage{pdflscape}	% Landscape pages
\usepackage[normalem]{ulem}
\newcommand{\specialcell}[2][c]{%
	\begin{tabular}[#1]{@{}c@{}}#2\end{tabular}}

\usepackage{multirow}
\newcommand{\kms}{\,km\,s$^{-1}$} % kilometres per second
\newcommand{\Hii}{H~{\sc ii}}

\title[Star formation timescale in WB~673]{Star formation timescale in the molecular filament WB~673}

\author[O. L. Ryabukhina et. al.]{O. L. Ryabukhina$^1$\thanks{Contact e-mail: \href{mailto:ryabukhina@ipfran.ru}{ryabukhina@ipfran.ru}}, M. S. Kirsanova$^1$, C. Henkel$^{2,3,4}$, D.~S. Wiebe$^1$
\\
$^1$Institute of Astronomy, Russian Academy of Sciences, 119017, 48 Pyatnitskaya Str., Moscow, Russia\\
	$^2$Max Planck Instit\"ut f\"ur Radioastronomie, Auf dem Hugel 69, 53121 Bonn, Germany \\
	$^3$Astronomy Department,King Abdulaziz University, PO Box 80203, 21589 Jeddah, Saudi Arabia \\
	$^4$Xinjiang Astronomical Observatory,    Chinese Academy of Sciences, 830011,    Urumqi, China}

\date{ }

\pubyear{2020}

\begin{document}
\label{firstpage}
\pagerange{\pageref{firstpage}--\pageref{lastpage}}
\maketitle

\begin{abstract}
  
We present observations of ammonia emission lines toward the interstellar filament WB~673 hosting the dense clumps WB~673, WB~668, S233-IR and G173.57+2.43. LTE analysis of the lines allows us to estimate gas kinetic temperature ($\lesssim$ 30~K in all the clumps), number density ($7-17\times10^3$~cm$^{-3}$), and ammonia column density ($\approx 1-1.5\times 10^{15}$~cm$^{-2}$) in the dense clumps. We find signatures of collapse in WB 673 and presence of compact spatially unresolved dense  clumps in S233-IR. We reconstruct 1D density and temperature distributions in the clumps and estimate their ages using astrochemical modelling. Considering CO, CS, NH$_3$ and N$_2$H$^+$ molecules (plus HCN and HNC for WB~673), we find a chemical age of  $t_{\rm chem}=1-3\times 10^5$~yrs providing the best agreement between the simulated and observed column densities in all the clumps. Therefore, we consider $t_{\rm chem}$ as the chemical age of the entire filament. A long preceding low-density stage of gas accumulation in the astrochemical model would break the agreement between the simulated and observed column densities. We suggest that rapid star formation over a $\sim 10^5$~yrs timescale take place in the filament.

\end{abstract}


%%%%%%%%%%%%%%%%%%%%%%%%%%%%%%%%%%%%%%%%%%%%%%%%%%

%%%%%%%%%%%%%%%%% BODY OF PAPER %%%%%%%%%%%%%%%%%%
\section{Introduction}

Interstellar molecular clouds appear as interconnected networks of elongated filaments \citep[e.~g.][and many others]{2009ApJ...700.1609M, Andre2014, 2016A&A...591A...5L}. According to numerical simulations, multiple episodes of interstellar gas compression by supersonic waves or shells produce these filaments \citep[see e.~g.][]{2009ApJ...704..161I, 2011ApJ...731...13N, 2015A&A...580A..49I, 2018PASJ...70S..53I}, but other scenarios, such as the development of hydrodynamic instabilities \citep[e.~g.][]{2001ApJ...553..227P} or cloud-cloud collisions \citep{2013ApJ...774L..31I, 2020A&A...642A..87K, 2021PASJ...73S...1F} are also possible. Particularly, \cite{2011ApJ...731...13N} show how the collision of two expanding gas shells, created by stellar winds or by supernova explosions, forming so-called bubbles, can lead to the creation of massive filamentary clouds (up to 10$^4$ M$_\odot$) in the region where the two bubbles overlap. Dense molecular clumps, being potential star formation sites, are formed after fragmentation of the filaments \citep[e.~g.][]{2000MNRAS.311..105F}. A high density of young stellar clusters and massive young stellar objects is observed in the clumps around infrared bubbles \citep[][]{Thompson_2012, 2014MNRAS.439.3719C, 2015ApJ...813...25S, Kendrew_2016}. \citet{2012ApJ...755...71K} estimated that the formation of up to 20\% of all massive stars in the Galaxy could be triggered  by the expanding H~II regions. Therefore, feedback from massive stars is an important ingredient that determines the star formation rate in molecular clouds. \cite{2018MNRAS.475.3511G} concluded that star formation is feedback-moderated in simulated turbulent molecular clouds over spatial scales from several to hundreds of parsecs.

Over the last twenty years, the majority of studies provide evidence in favour of a `rapid' star formation scenario with a typical timescale of $\sim 10^5-10^6$~yrs. Namely, \cite{1999ApJS..123..233L} estimate a typical lifetime of starless cores as $0.3-1.6\times10^6$~yrs using a sample of 406 dense cores. \cite{2003ApJ...585..850M} simulate the formation of massive stars in dense clumps and suggest a shorter timescale for the formation of a massive star, namely  $\sim 10^5$~yrs. \citet{2007ApJ...657..870V} also advocate rapid star formation over $\sim 10^5$~yrs in the clumps, albeit with a long (up to several Myrs) period of initial gas accumulation into the cores. \cite{2000MNRAS.311...63J} and \cite{2007prpl.conf...33W} estimate a lifetime of $\sim10^6$~yrs for cold cores without active star formation. \citet{2021A&A...646A.170G} report a rapid formation of the Serpens filament during less than $2\times10^6$~yrs using a timescale for the CO freeze-out process. They showed that the chemical timescale agrees with the age of this filament, obtained with a time-dependent accretion model.

The aim of the present study is to estimate a chemical age of the molecular filament WB~673 using a set of molecular abundances including both `early' (CO, CS) and `late' (NH$_3$, N$_2$H$^+$) species, with NH$_3$ data presented in this paper and other abundances taken from \citet{2020ARep...64..394R}. We confirm the scenario of the `rapid' formation for dense clumps in high-mass star-forming regions without a long initial stage of gas accumulation in the WB~673 filament, by showing that the chemical age of the filament is $1-3\times10^5$~yrs. We also show that ammonia is depleted in the dense clumps. In addition to CO and CS depletion, found by \citet{2020ARep...64..394R}, this indicates that internal heating sources, related to embedded high-mass young stellar objects, still do not impact much chemistry of the surrounding medium.

\section{TARGET REGION}

The giant molecular cloud G174+2.5 at a distance of 1.6 kpc (\citet{2015MNRAS.453.3163B}; $40\arcsec$ correspond to 0.3 pc) was extensively mapped in CO lines by \cite{1996ApJ...463..630H} and \cite{2016ApJS..226...13B}. They found that the spatial distribution and kinematics of the molecular gas are affected by the extended and optically prominent \Hii{} regions Sh2-231, Sh2-232, Sh2-235 (S231, S232 and S235 hereafter) around massive O-type stars. There are several B-type stars in G174+2.5, which are deeply embedded in parental molecular material. They are surrounded by the compact \Hii{} regions S233, S235A and S235C. A significant fraction of the molecular gas in G174+2.5 is concentrated in a single filamentary cloud, WB~673, with a projected length of 25~pc  ($\sim$50 arcmin), total mass of $10^{4}$~M$_\odot$ and mass-to-length ratio of $3.4-34$~M$_\odot$~pc$^{-1}$ \citep{2017OAst...26...99K}. The filament is located between S231 and a large unnamed envelope, which can be a supernova remnant \citep[][]{2012AJ....143...75K,2017OAst...26...99K}. Therefore, WB~673 is a reliable candidate of a region exposed to multiple compression events from expanding shells.

\begin{figure}
	\includegraphics[width=\linewidth]{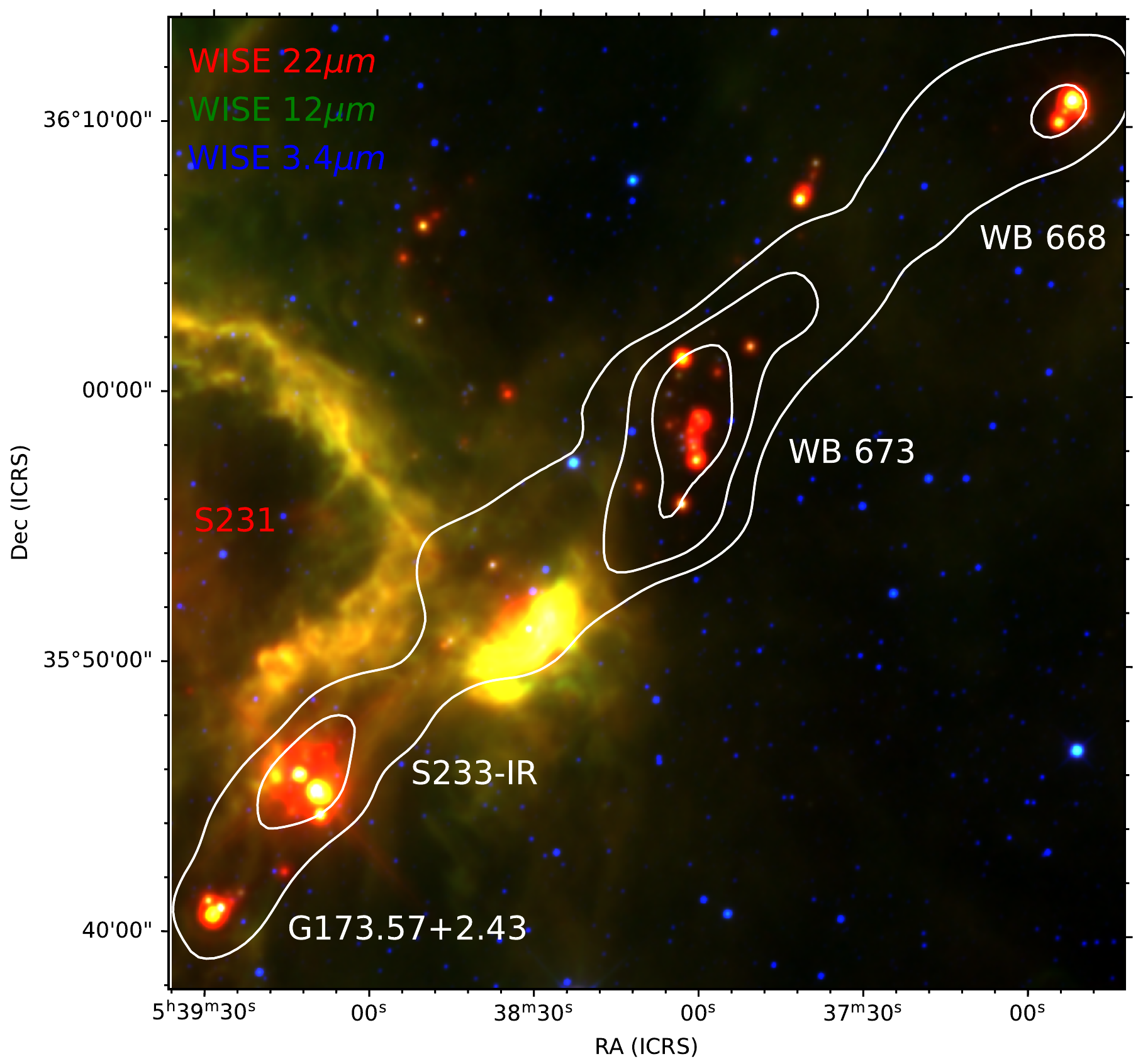} 
	\caption{Pseudo-colour image of the WB~673 filament composed of WISE IR data \citep{2010AJ....140.1868W} at 22~$\mu$m (red), 12~$\mu$m (green), and 3.4~$\mu$m (blue). The integrated intensity of the $^{13}$CO (1--0) emission line is shown by white contours at levels of 14, 24, 35 K~\kms from \citet{2020ARep...64..394R}; beam size is $43\arcsec$. An arch, related with S231, is seen to the east of the filament.}
	\label{fig_wise}
\end{figure}

\citet{2016AstBu..71..208L} and \citet{2017OAst...26...99K} observed emission of various molecular lines at wavelengths $\lambda\sim3$ and 1.3~mm in four dense clumps within the WB~673 filament. A mid-infrared image of the filament and a map of the $^{13}$CO (1--0) line emission are shown in Fig. \ref{fig_wise}, where the dense clumps are designated as WB~668, WB~673 \citep[WB89~668 and WB89~673 in][]{1989A&AS...80..149W}, S233-IR and G173.57+2.43. Water masers at 22 GHz, Class~II methanol masers and IRAS (InfraRed Astronomical Satellite) and MSX (Midcourse Space Experiment) sources were found in all the clumps, which is a sign of active star formation \citep[see][for details]{2016AstBu..71..208L}. The infrared flux ratios in IRAS point sources 05345+3556 and 05358+3543 are consistent with \Hii{} regions in WB~673 and S233-IR. The latter clump hosts a well-studied molecular outflow \citep[e.~g.][]{2007A&A...466.1065B, 2007A&A...475..925L}.

\citet{2020ARep...64..394R} found that  abundances  of CS, CO, N$_2$H$^+$, HCN and HNC  relative to H$_2$ decrease in the dense central parts of the clumps in comparison with their periphery. In other words, the clumps demonstrate typical cold chemistry in spite of signs of active high-mass star formation. In the following we investigate the filament where the star formation process is at its early stage.

\section{Observations}\label{sec_obs}

We performed observations of the ammonia (1,1), (2,2) and (3,3) inversion transitions with the 100-m Effelsberg telescope (Germany) in 2--7 January 2019. The total ON+OFF observational time was 60~hours. The observations were carried out in continuous mapping mode (on-the-fly) using  a 1.3~cm secondary focus receiver with a bandwidth of 300~MHz, providing a spectral resolution of 0.2~\kms. The ammonia emission data covered the entire area of the filament with a size $10\arcmin\times 50\arcmin$ with respect to  the minor and major axes. The obtained maps have an angular resolution of 40\arcsec. The observations were done in the position switching mode with the OFF-position at $\alpha$(ICRS)~=~05$^{\rm h}$37$^{\rm m}$00$^{\rm s}$, $\delta$(ICRS)~=~+35$^\circ$30$^\prime$00$^{\prime\prime}$. The list of observed transitions is given in Table~\ref{table:lines}.

The calibration involved a correction for the atmospheric opacity, that is based on an atmospheric model and the water vapour radiometer at the Effelsberg telescope. The elevation dependent telescope gain was also implemented. The antenna efficiencies and sensitivities (i.e. Kelvin per Jansky coefficients) were derived from an analytical model and the full width at half maximum (FWHM) of the pointing measurements. The temperature of the noise diode, providing an initial intensity scale, was calibrated using radio continuum pointing scans toward the planetary nebula NGC 7027 \citep{1994A&A...284..331O}.
%with coordinates $\alpha$(ICRS)~=~21$^h$07$^m$1.752$^s$, $\delta$(ICRS)~=~+42$^\circ$14$^\prime$9.96$^{\prime\prime}$. 
Pointing and focus were performed approximately every hour. $T_{\rm sys}$ was mostly kept within the range of 100--120 K on a $T_{\rm A^*}$ scale. However, it increased to 200~K in bad weather conditions.  We excluded spectra with a noise level higher than $\approx 0.3$~K from our analysis. 
The median of the signal-to-noise ratio of the determined lines is three. In the densest regions at the centers of the WB~673 and S233-IR clumps it exceeds 10 and 5 for the (1,1) and (2,2) transitions, respectively.
% The average uncertainty of the determined line intensities is 7\% and 13\% for the (1,1) and (2,2) transitions, respectively.

The gridding of the data and the baseline correction were processed using the CLASS program from the GILDAS\footnote{http://www.iram.fr/IRAMFR/GILDAS} \citep{Maret2011} package. Further analysis was done using MIRIAD \citep{Sault1995}, Astropy \citep{2018AJ....156..123A} and PySpecKit \citep{2011ascl.soft09001G} packages. 

\begin{table}
\centering
	\caption{  \textbf{List of observed transitions}}
	\label{table:lines}
	\bigskip
	\begin{tabular}{cccc}
		\hline
		Molecule & Transition  &  Frequency & E$_u$  \\
		&& (GHz) &(K)\\
		\hline
		 & (1,1) & 23.6945 & 23.4 \\
		NH$_3$ & (2,2) & 23.7226 & 64.9 \\
		 & (3,3) &  23.8701 & 124.5  \\
		\hline
	\end{tabular}
\end{table}

\section{Data analysis}\label{sec_analysis}

The lowest meta-stable energy states of ammonia are only excited by collisions. Therefore they are widely used for measuring the gas temperature in dense molecular gas. A review and a theory are presented in \citet{1983ARA&A..21..239H}. Our LTE analysis of the ammonia lines follows this classical paper. 

We used the CLASS routine 'nh3(1,1)' to estimate the optical depth of the $(J,K)=(1,1)$ line, connecting the 
        two states of the $(J,K)=(1,1)$ inversion doublet,
        accounting for hyperfine components and excitation 
        temperature  $T_{\rm ex}$. This method assumes Gaussian line shapes in the optically thin case and equal excitation temperatures for all the hyperfine components. For the (1,1) transition, the optical depth of the main (the central one of five) group of    hyperfine components  $\tau_{(1,1)\mathrm m}$ is related to the total optical depth over the multiplet given by the 'nh3(1,1)' routine as $\tau_{(1,1)\mathrm m}$ = $\tau_{(1,1)}$/2 (see Appendix in \citet{1992ApJ...388..467M}). Moreover, we found the intensity of the main components ${T}_{\rm MB}$, LSR velocity $V_{\rm LSR}$ and line width $\triangle V$ using PySpecKit for all three observed transitions. Their values for the peaks of the clumps are given in Table~\ref{peak_par}).

Using the (1,1) and (2,2) line intensities, we can find the rotational temperature, which is consistent with the relative populations of these two inversion 
         doublets:
\begin{equation}\label{eq_Trot_tau}
	T_{\rm rot} = -41.5\ln\left(\frac{-0.282}{\tau_{(1,1)\mathrm m}} \ln\left(1 - \frac{{T}_{\rm MB(2,2)}}{T_{\rm MB(1,1)}}(1 - \exp(-\tau_{(1,1)\mathrm m})) \right)\right)^{-1} (\mathrm K),
\end{equation}
where $T_{\rm MB(1,1)}$ and $T_{\rm MB(2,2)}$ are the main beam brightness temperatures of the main components of the (1,1) and (2,2) lines, respectively. The ammonia column density at the (1,1) level can be calculated
  following \citet{1992ApJ...388..467M} with:
\begin{equation}
N_{1,1} = 6.6 \times 10^{14}\frac{T_{\rm rot}}{\nu_{(1,1)}}\tau_{(1,1)m}  \triangle V_{(1,1)} (\rm cm^{-2}),
\end{equation}\label{eq1}
where $\nu_{(1,1)}$ is the frequency of the NH$_3$~(1,1) in GHz and $\triangle V_{(1,1)}$ is the line width in kilometres per second. In those pixels, where the (2,2) peak intensity is at a $<3\sigma$ level, we assumed $T_{\rm rot}=10$~K, to be consistent with \cite{2020ARep...64..394R}.

In those directions, where $\tau_{(1,1)\rm m}\ll$ 1 or the hyperfine components are not visible, the (1,1) line is considered as optically thin. In this case $T_{\rm rot}$ and $N_{1,1}$ are determined in the low optical depth approximation:
\begin{equation}\label{eq_Trot_int}
T_{\rm rot} = -41.5/\ln\left(0.2\frac{\int {\textit T}_{\rm MB(2,2)}{\rm d}V}{\int {\textit T}_{\rm MB(1,1)}{\rm d}V}\right) (\mathrm K),
\end{equation}
where $\int {\textit T}_{\rm MB(2,2)}dv$ and $\int {\textit T}_{\rm MB(1,1)}dv$ are the integrated intensities of the (2,2) and (1,1) lines, respectively (see Appendix).  The $N_{1,1}$ value is determined as:
\begin{equation}
N_{1,1} = 3.3 \times 10^{14} \frac{T_{\rm rot}}{\nu_{(1,1)}} \frac{\int T_{\rm MB(1,1)} dV}{J(T_{\rm rot}) - J(T_{\rm bg})}(\rm cm^{-2}),
\end{equation}
where $T_{\rm bg}$ = 2.73~K is the background temperature, and $J(T) = \left( h \nu_{(1,1)} / {\rm k} \right) / \left( \exp\left({\frac{h \nu_{(1,1)}}{{\rm k} T}}\right) - 1 \right)$

We determined the total ammonia column density over the lowest metastable energy levels as:
\begin{equation}
\begin{gathered}
N_{\rm NH_3} = N_{1,1}\bigg(\frac{\exp({21.3/T_{\rm rot}}) }{3} + 1  +  \frac{5 \exp({-41.2/T_{\rm rot}})  }{3} + \\ + \frac{14 \exp({-99.4/T_{\rm rot}})}{3}\bigg)(\rm cm^{-2}).
\end{gathered}
\end{equation}

\citet{1983A&A...122..164W} and \citet{2004A&A...416..191T} provide an equation for the gas kinetic temperature ($T_{\rm kin}$):
\begin{equation}
T_{\rm kin} = \frac{T_{\rm rot} }{1 - \frac{T_{\rm rot}}{41.5}\ln(1 + 1.1\exp(\frac{-16}{T_{\rm rot} }))} (\rm K).
\end{equation}
where 41.5 K is the energy gap between the (1,1) and (2,2) levels. This equation is derived from fitting $T_{\rm kin}$ to the density distribution $n(r) = n_0/(1 + (r/r_0)^{2.5})$ to compare their observationally determined approximately constant rotational temperatures with modeled kinetic temperatures in dense quiescent molecular clouds \citep{2004A&A...416..191T, 2022A&A...658A..34T}.

Using the obtained $T_{\rm ex}$ and $T_{\rm kin}$ values, \cite{1983ARA&A..21..239H} gave the following equation for the gas number density:
\begin{equation}
n({\rm H_2}) = \frac{A}{C}\frac{J(T_{\rm ex}) - J(T_{\rm bg})}{J(T_{\rm kin}) - J(T_{\rm ex})}\left[ 1 + \frac{J(T_{\rm kin)}}{{h} \nu / {k}}\right] ({\rm cm^{-3}}),
\end{equation}
where $A$ and $C$ are the Einstein coefficient for spontaneous emission and the collision rate, respectively. According to \cite{2005A&A...432..369S}, for typical $T_{\rm kin}$ = 25 K, the coefficients become $A = 1.7\times10^{-7}$~s$^{-1}$ and $C = 8.6\times10^{-11}$~cm$^3$s$^{-1}$.

Under LTE conditions and without a line-of-sight velocity gradient, the two inner satellite lines and two outer satellite lines of the NH$_3$~(1,1) transition have the same pairwise intensities. However, sometimes hyperfine intensity anomalies are observed in molecular clouds \citep[see e. g.][]{1985A&A...144...13S, 2001A&A...376..348P, 2020A&A...640A.114Z}, in the sense that the intensities of red-shifted components are not equal to the intensities of their blue-shifted counterparts. We  quantify these anomalies calculating the ratios of the satellite's intensities after fitting them by independent Gaussian functions.  We designate intensities of the red-shifted inner (I) and outer (O) satellites divided by intensities of the corresponding blue-shifted satellites as $A_{\rm IS}$ and $A_{\rm OS}$. The ratios $A_{\rm OS}=A_{\rm IS}=1$ imply no anomaly. The case of $A_{\rm OS}>1$ and $A_{\rm IS}<1$ (brighter red-shifted outer component and fainter red-shifted inner component) can be explained by a model of unresolved small dense gas condensations embedded into a less dense medium \citep[see][]{1985A&A...144...13S}. Local velocity gradients can explain the case of $A_{\rm OS}>1, A_{\rm IS}>1$ (expansion) or $A_{\rm OS}<1, A_{\rm IS}<1$ (contraction) \citep[see e.g. the radiative transfer calculations of][]{2001A&A...376..348P, 2004R&QE...47...77T, Pavyar_2008}.

\section{Results of the observations}

\subsection{Ammonia emission in the dense clumps}

\begin{figure*}
	\includegraphics[width=\linewidth]{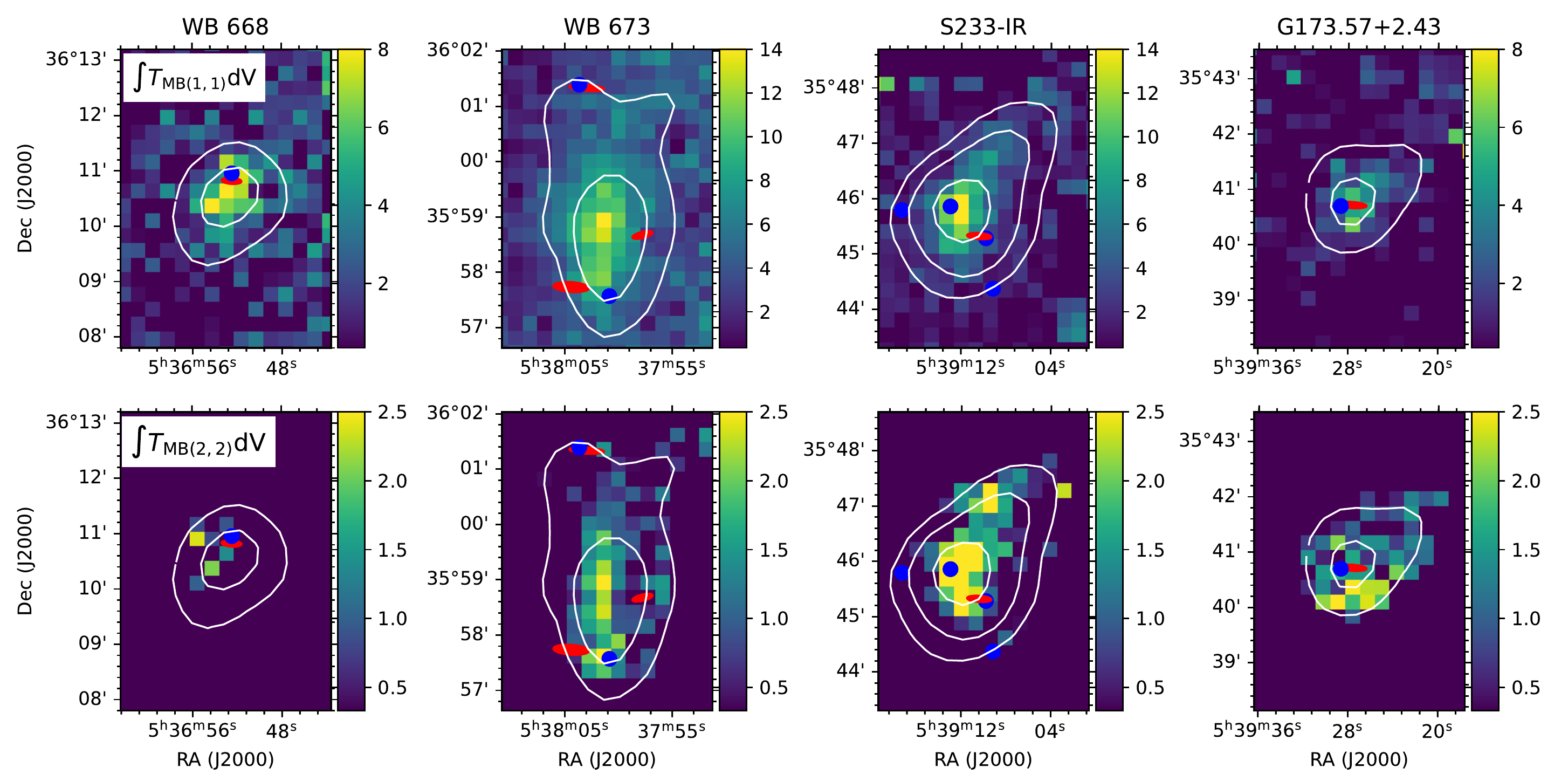} 
	\caption{Integrated intensities of NH$_3$~(1,1) (top) and NH$_3$~(2,2) (bottom) lines from WB~668, WB~673, S233-IR and G173.57+2.43. The colour bar shows the intensities in K~km~s$^{-1}$. Maps are cut at levels of 3$\sigma$, which are 0.35 and 0.33 K~km~s$^{-1}$ for the (1,1) and (2,2) lines, respectively. Compact red ellipses show IRAS sources (the size of the ellipse corresponds to the position uncertainties), compact blue circles show MSX sources. White contours provide H$_2$ column density, where the contours correspond to $7 \cdot 10^{21}, 2 \cdot 10^{22}$ and $7 \cdot 10^{22}$ cm$^{-2}$ \citep{2020ARep...64..394R}.}
	\label{fig_int11}
\end{figure*}

Maps of the integrated intensities of the NH$_3$~(1,1) and (2,2) lines in the dense clumps WB~668, WB~673, S233-IR and G173.57+2.43 are shown in Fig.~\ref{fig_int11}. The critical density of the observed ammonia lines is quite high: $n_{\rm crit} = 1.8 \times 10^3$ cm$^{-3}$ for $T_{\rm kin} = 20$~K  \citep{2015PASP..127..299S}. Therefore, in the absence of depletion onto dust, the ammonia emission peaks mark the densest regions of the clumps. 
There is a coincidence between the ammonia emission peaks and IRAS sources in all the observed clumps except WB 673, where the sources are situated at the periphery of the clump with $N_{{\rm H}_2}< 2 \cdot 10^{22}$ cm$^{-2}$. However, all positions of the ammonia peaks correlate with the dust emission peaks  of the 1.1 mm Bolocam data. While we do not have a convincing explanation for this discrepancy, in the following we will use the Bolocam data, because they were obtained at longer wavelengths, thus sampling the entire dust. To summarize, the dust emission peaks do not only indicate high column densities but are also characterised by particularly high volume densities. Contours in Fig.~\ref{fig_int11} show the H$_2$ column densities ($N_{{\rm H}_2}$), with the levels being $7 \cdot 10^{21}$, $2 \cdot 10^{22}$, and $7 \cdot 10^{22}$ cm$^{-2}$ \citep{2020ARep...64..394R}. The NH$_3$~(1,1) integrated intensity increases with total hydrogen column density. The brightest lines appear in the WB~673 and S233-IR clumps, where the peak intensities reach 18~K~\kms, and 5~K~\kms for the (1,1) and (2,2) lines, respectively. Weaker ammonia lines are found in the clumps WB~668 and G173.57+2.43, where the peak intensities reach 8~K~\kms\ for the (1,1) transition and 2.5~K~\kms\ for the (2,2) transition. 

\begin{figure*}
	\includegraphics[width=0.99\linewidth]{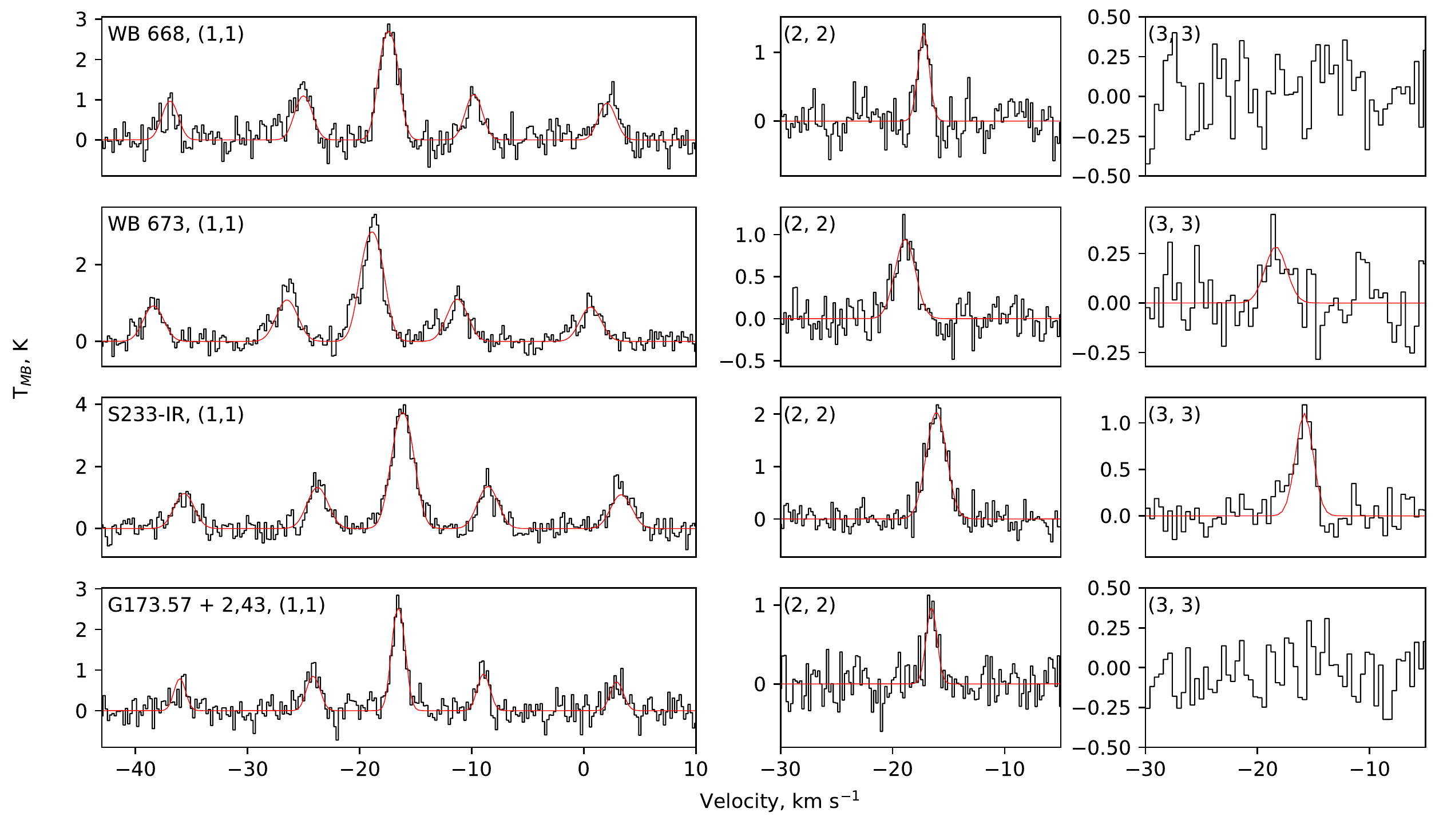} 
	\caption{The observed ammonia lines at the emission peaks. Red lines show Gaussian fits obtained with the CLASS Gildas package.}
	\label{fig_spectra}
\end{figure*}

Examples of the NH$_3$~(1,1), (2,2) and (3,3) lines at the ammonia emission peaks are shown in Fig.~\ref{fig_spectra}. From the initial spectral resolution of 0.2~\kms, the spectra of the (3,3) transitions are smoothed to 0.4~\kms. The five groups of  hyperfine components are detected in the (1,1) line towards dense clumps.   Regions between these clumps are often devoid of 
     detected emission.          Parameters of the Gaussian fits are given in Table \ref{peak_par}, where $\tau_{(1,1)}$ is the total optical depth over the (1,1) multiplet, $T_{\mathrm MB}$ and $\triangle V$ are the main beam brightness temperature and the width of the (1,1), (2,2) and (3,3) lines. In all the observed       positions the line widths contain a thermal and a non-thermal contribution. The thermal line width is 0.22~\kms for a gas temperature $T = 20$~K (see below), while the observed line widths are $\approx 1.1$~\kms in G173.57+2.43 and $\approx 2.0$~\kms in the other clumps.

We find a moderate peak optical depth of the ammonia emission $\tau_{(1,1)} < 2$ in all observed clumps, where $\tau_{(1,1)} \approx 0.8$ in the two inner clumps (WB~673 and S233-IR) and reaching 1.8 in the two outer clumps (WB~668 and G173.57+2.43). 

\begin{table*}
	\caption{\textbf{Line parameters at the peaks of the NH$_3$ (1,1) integrated intensity.}}
	\label{peak_par}
	\bigskip
	\begin{tabular}{ccccccccccc}
		\hline
		Clump & \specialcell{$\alpha_{2000}$\\ h m s}  &  \specialcell{$\delta_{2000}$\\ $\circ $ $ \prime$ $\prime\prime$ } & $\tau_{(1,1)}$ & \specialcell{$T_{\rm MB(1,1)}$, \\ K} &  \specialcell{$\triangle V_{(1,1)}$, \\  \kms} &  \specialcell{$T_{\rm MB(2,2)}$, \\ K} &  \specialcell{$\triangle V_{(2,2)}$, \\ \kms} & \specialcell{$T_{\rm MB(3,3)}$, \\ K}& \specialcell{$\triangle V_{(3,3)}$, \\ \kms} & \\ \hline
		WB~668  & 5 36 53 & +36 10 37 & 
		
		1.38  $\pm$ 0.29 & 
		2.71 $\pm$ 0.30& 1.62 $\pm$ 0.07 
		& 1.29 $\pm$ 0.26 & 1.11 $\pm$ 0.14
		& - & - \\
		
		WB~673  & 5 38 01 & +35 58 55 
		& 1.05 $\pm$ 0.17 
		& 2.85 $\pm$ 0.18 	& 2.04 $\pm$  0.09 	
		& 0.94 $\pm$ 0.16 & 2.08 $\pm$ 0.18
		& 0.28 $\pm$ 0.11 & 2.17 $\pm$ 0.61 \\
		
		S233-IR  & 5 39 12 & +35 45 53 
		& 0.83 $\pm$ 0.16 
		& 3.74 $\pm$ 0.24 	& 2.03 $\pm$ 0.05	
		& 2.02 $\pm$ 0.22 & 2.07 $\pm$ 0.10
		& 1.10 $\pm$ 0.16& 1.74 $\pm$ 0.20 \\
		
		G173.57+2.43 & 5 39 27 & +35 40 35 
		& 0.84 $\pm$ 0.31 
		& 2.52 $\pm$ 0.20	& 1.07 $\pm$ 0.20 	
		& 0.96 $\pm$ 0.21	& 1.92 $\pm$ 0.39
		& - & - \\
		[1mm]
		\hline
	\end{tabular}
\end{table*}

\subsection{Physical conditions in the dense clumps}\label{sec_phys_cond}

\begin{figure*}
	\includegraphics[width=1\linewidth]{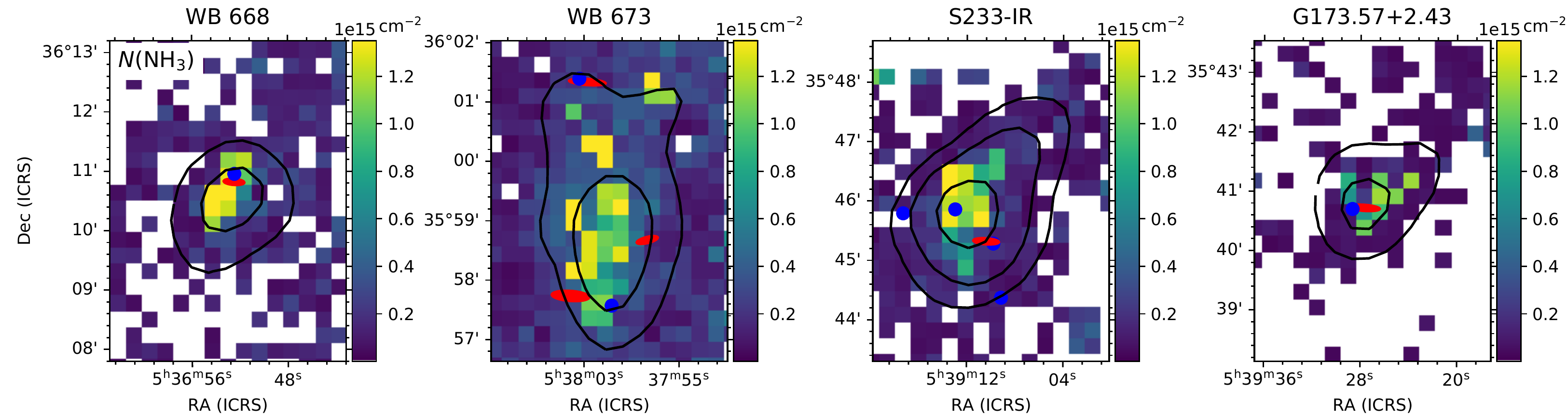} 
	\includegraphics[width=1\linewidth]{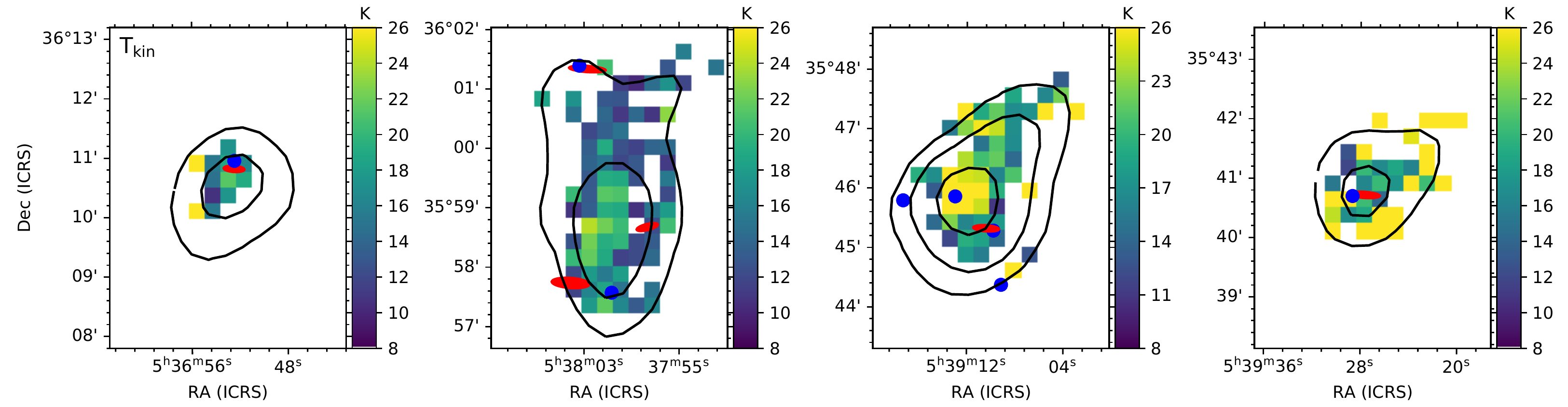} 
	\includegraphics[width=1\linewidth]{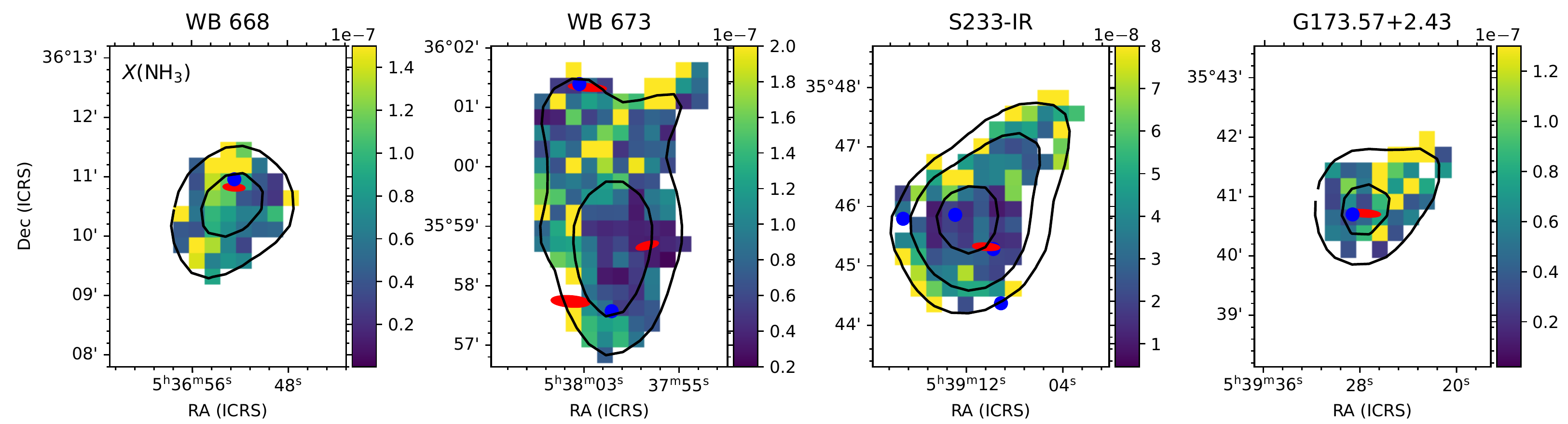} 
	\includegraphics[width=1\linewidth]{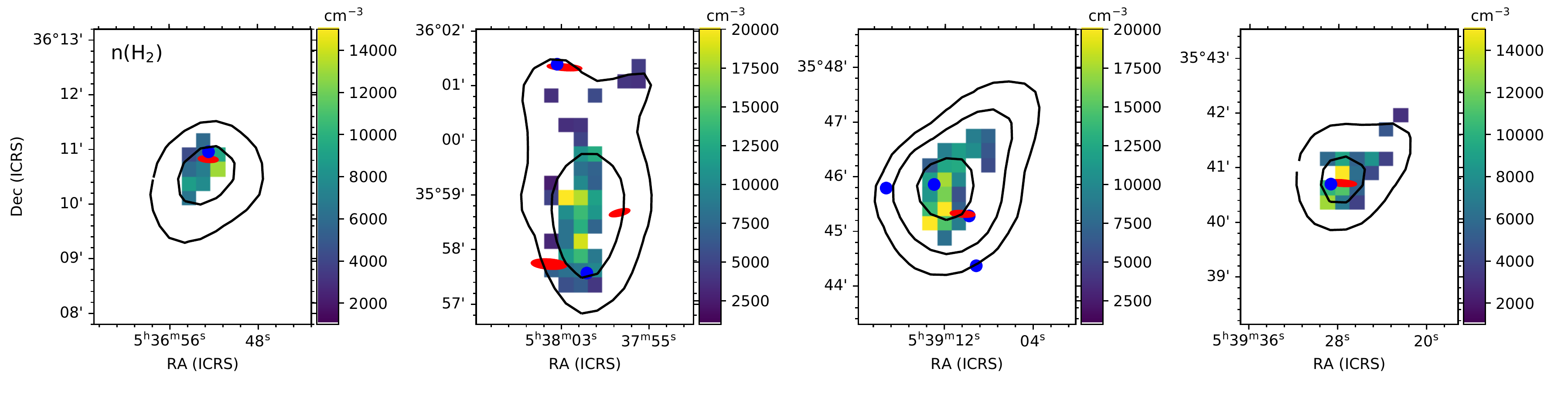} 
	\caption{Maps of the ammonia column density (top), gas kinetic temperature (middle top), relative NH$_3$ abundance (middle bottom) and $n({\rm H2})$ (bottom) in the observed clumps. Red ellipses show IRAS sources, blue circles show MSX sources. Black contours provide H$_2$ column density, where the contours correspond to $7 \cdot 10^{21}, 2 \cdot 10^{22}, 7 \cdot 10^{22}$ cm$^{-2}$ \citep{2020ARep...64..394R}.}
	\label{fig_N}
\end{figure*}

Maps of the ammonia column density ($N_{\rm NH_3}$), gas kinetic temperature ($T_{\rm kin}$) and the relative ammonia abundance ($X_{\rm NH_3}$) in the dense clumps are presented in Fig.~\ref{fig_N}. The highest $N_{\rm NH_3}$ values coincide with the dust emission peaks at 1.1~mm as we expected from the integrated intensity maps. Peak values are $N_{\rm NH_3} = 1-2\times 10^{15}$~cm$^{-2}$ in WB~668, WB~673 and S233-IR, but the value is lower by a factor of two in G173.57+2.43. 

The gas kinetic temperature is confined to the  range of 18--30~K. The highest $T_{\rm kin} \approx 30$~K values are found in the central part of S233-IR. Values of $T_{\rm kin} \approx 18-22$~K are observed throughout the whole area of WB~668 and WB~673. In G173.57+2.43, the inner part with $T_{\rm kin} \approx 18$~K is surrounded by a warmer envelope with $T_{\rm kin} \approx 26$~K.  Therefore, these four observed clumps represent several distinct environments, namely a warm core with a cold envelope (S233-IR), a cold core with a warm envelope  (G173.57+2.43) and quite uniform temperature distributions in WB~668 and WB~673. Median relative temperature errors are 12\% for the WB~668 clump, 6\% for WB~673, 10\% for S233-IR and 16\% for G173.57+2.43

The relative ammonia abundance $X_{\rm NH_3}$ decreases by up to an order of magnitude in the central and densest parts of WB~673 and S233-IR relative to their outskirts. The maximum $X_{\rm NH_3} = 2\times 10^{-7}$ is observed in the northern part of WB~673, but decreases to $2\times10^{-8}$ at the dust emission peak. The minimum abundance $\approx 1-2\times 10^{-8}$ is found in the centre of S233-IR, and the abundance reaches $8\times10^{-8}$ at the periphery of the clump. There are north-south and west-east abundance gradients from about 0.2 to $1\times 10^{-7}$ in the WB~668 and G173.57+2.43 clumps, respectively, but no central depletion of ammonia abundance is discernible in the cores of these clumps. We note that the lowest value of  $X_{\rm NH_3}$ is observed in the warmest clump S233-IR. There are no other trends between such parameters as $N_{\rm NH_3}$, $T_{\rm kin}$ and $X_{\rm NH_3}$ in the observed clumps, probably due to their complex structure unresolved in our data. Physical parameter values for the dust emission peaks at 1.1~mm with their respective errors are given in Table~\ref{table:plummer_par}.

The gas number density reaches $2\times 10^4$~cm$^{-3}$ in the two inner clumps (WB~673 and S233-IR), while it is $1.5 - 2$ times less in the outer clumps. The density peak in WB~673 coincides with the maximum of the dust continuum emission, while it is shifted to the south by $\sim36''$ relative to the 
         dust peak in S233-IR. The gas density peaks coincide with the dust emission peaks and locations of the infrared sources in the two peripheral clumps.

\begin{figure}
	\includegraphics[width=1\linewidth]{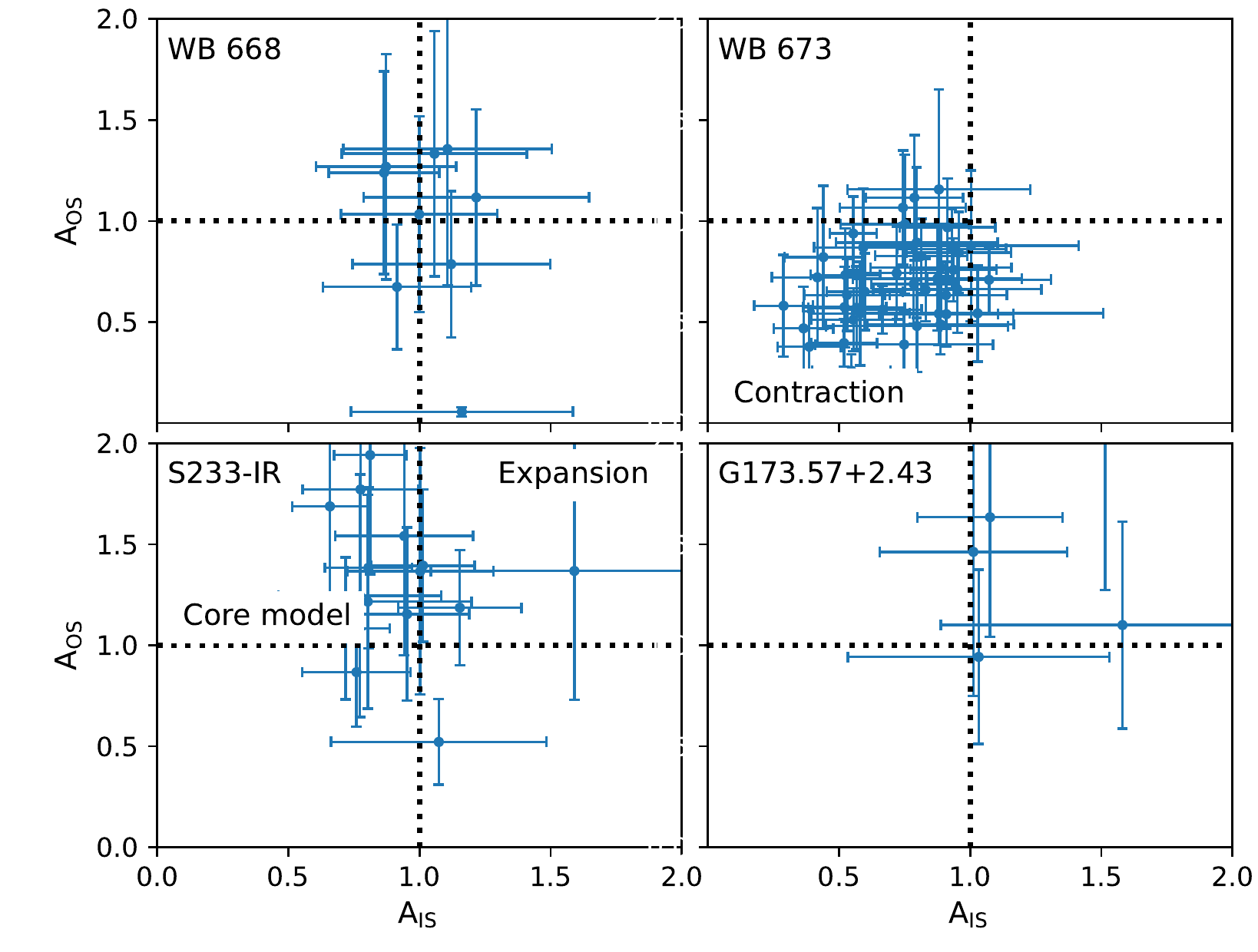} 
	\caption{Various models of hyperfine intensity anomalies in the dense clumps (see the end of Sect. \ref{sec_analysis}) .}
	\label{fig_corr_hia}
\end{figure}

Fig. \ref{fig_corr_hia} allows us to analyse the hyperfine intensity anomalies using $A_{\rm IS}$ and $A_{\rm OS}$ values. The dotted lines divide the plot into separate quadrants, where the values pertaining to the various situations mentioned above (see end of Sect. \ref{sec_analysis}) reside. 
Most $A_{\rm IS}+A_{\rm OS}$ pairs in the WB~673 clump match the local velocity gradient model corresponding to an infall motion. In the S233-IR clump, most pairs fall into the upper left quadrant with $A_{\rm IS} < 1$ and $A_{\rm OS}>1$, which corresponds to the model of small
     unresolved clumps. There are also several pairs
     in the upper right quadrant weakly hinting at 
     an expanding motion. No significant hyperfine
     anomalies are seen in WB668 and G183.57+2.43.

\section{Astrochemical modelling}\label{chapt:mod}

In order to estimate the chemical ages of the observed clumps, we performed astrochemical modelling relying upon the physical parameters found in the present study. For that purpose, we used the Presta astrochemical model \citep[see description of the model in][]{2013ARep...57..818K}, which includes both gas-phase and solid-phase processes taken from the ALCHEMIC database \citep{alchemic} with additions described by \cite{wiebe2019}. The initial gas chemical composition was taken from \citet{1996A&AS..119..111L} and is presented in Table \ref{table:init_abud}. We start with purely atomic initial abundances and follow the chemical evolution for $10^8$~years.

\begin{table}
\centering
	\caption{  \textbf{Initial fractional abundances with respect to the total hydrogen abundance}}
	\label{table:init_abud}
	\bigskip
	\begin{tabular}{cc}
		\hline
		Element & Abundance \\ \hline 
		He & $9.0 \cdot 10^{-2}$ \\
		C & $7.3 \cdot 10^{-5}$ \\
		N & $2.1 \cdot 10^{-5}$ \\
		O & $1.7 \cdot 10^{-4}$ \\
		Cl & $1.0 \cdot 10^{-9}$ \\
		S & $8.0 \cdot 10^{-8}$ \\
		Si & $8.0 \cdot 10^{-9}$ \\
		Fe & $3.0 \cdot 10^{-9}$ \\
		\hline
	\end{tabular}
\end{table}

The Presta model simulates an evolution of a 1D object. In this study, we set up clump physical parameters assuming spherical symmetry. This is justified as all the considered cores appear as isolated dense objects surrounded by less dense envelopes in our observations. They probably contain more complex inner structure,
         but this is not resolved in our observations. We made radial cuts through the number density maps (Fig.~\ref{fig_N}, bottom row) and averaged the cuts over the position angle. After that, we fitted the averaged density profiles using a quasi-universal Plummer-like function,
\begin{equation}\label{eq_Plum}
n(r) = \frac{n_0}{(1 + (r / r_0)^2)^{p/2}},
\end{equation}
which is believed to be appropriate for the clumps belonging to the filamentary structure \citep{Andre2014}. Here $n_0$ is the density at the centre of the clump, $r_0$ is the radius of the flat inner region, $p$ is the power-law exponent at $r > r_0$. Flat inner region means that at a radius less than $r_0$, the density 
         is approxiamtely $n_0$. Derived profile parameters are given in  Table~\ref{table:plummer_par}. The density profile for clump WB~673 is shown in Figs.~\ref{fig_wb673_model}. Since the temperature range in the clumps is narrow, a constant temperature $T$ is adopted in all the models. Gas and dust temperatures are assumed to be equal. Figs.~\ref{fig_wb673_model} also shows radial profiles of the column densities of ammonia (this study) and other molecules \citep[N$_2$H$^+$, CS, CO, HCN, HNC, taken from][]{2020ARep...64..394R} averaged over the azimuthal angle. Error bars over the radial profiles show the scatter of original values.

\begin{table*}
\centering
	\caption{  \textbf{Parameters of the density distribution and derived physical parameters at clump centres}}
	\label{table:plummer_par}
	\bigskip
	\begin{tabular}{cccc|cccc}
		\hline
		\multirow{2}{*}{Clump} & \multicolumn{3}{|c|}{Plummer-like density distribution}   &  \multicolumn{3}{|c|}{Derived physical parameters}  \\
		& $n_0$ & $r_0$  & $p$ & $T_{\rm kin}$ & $N$(NH$_3$) & $x$(NH$_3$) & \\
		& $10^3$ cm$^{-3}$ & pc  & -- &  K & $10^{14}$ cm$^{-2}$ & $10^{-8}$ \\
		\hline
		WB~668  & 7.4 & 0.126 & 0.20 & 22 $\pm$ 2 & 12 $\pm$ 4 & 9 $\pm$ 3 \\
		WB~673  & 13.3 & 0.219 & 0.40 &  19 $\pm$ 1 & 9 $\pm$ 3 & 5 $\pm$ 1 & \\
		S233-IR & 16.9 & 0.014 & 0.11 & 29 $\pm$ 2 & 12 $\pm$ 3 & 3 $\pm$ 1 & \\
		G173.57+2.43 & 12.8 &  0.040 & 0.35 & 20 $\pm$ 2 & 7 $\pm$ 3 & 5 $\pm$ 2 & \\
		\hline
	\end{tabular}
\end{table*}

The Presta model produces radial abundance profiles for various molecules at different moments of the cloud evolution. These relative abundances are then multiplied by the hydrogen column density and integrated over the radial profile to obtain column densities.

In order to check a consistency between simulated $N_{\rm model}$ and observed $N_{\rm obs}$ molecular column densities, we introduce a correspondence criterion $\Sigma$ in the following way. If $\log|N_{\rm obs}/N_{\rm model}|\le1$ at all points along the clump projected radius, then the model and the observations are assumed to be in agreement, and $\Sigma=1$; otherwise $\Sigma$ = 0. Such criteria were evaluated for each considered molecule and then summed up, so that the maximum $\Sigma$ value is four for all the clumps except WB~673. In view of the HCN and HNC data from \citep{2020ARep...64..394R}, for WB~673 the maximum $\Sigma$ is~6.

\section{Modelling Results}

%Using the time-dependent astrochemical model and checking the criteria of correspondence at each model time step, we estimated a chemical age, at which the simulated and observed molecular column densities are in best agreement.

We used a chemical model with $T=20$\,K and zero intensity of the radiation field ($G_0=0$) as a reference model in our study. The adopted cosmic ray ionisation rate was $\zeta=3\times 10^{-17}$~s$^{-1}$ for each clump. The simulations were performed for a broad range of input parameters, varied relative to the reference model. Namely, we tried lower and higher temperatures (up to $\pm$ 10\,K), hydrogen volume density $n_0$ (up to $\pm$ one order of magnitude), cosmic ray ionisation rate (up to one order of magnitude). Also various non-zero interstellar radiation fields with depth-dependent $A_{\rm V}$ were tested. In addition, we considered a model with a long ($5 \cdot 10^5$~yrs) preceding stage of gas accumulation at $T=10$~K, the characteristic temperature for dark clouds (e.g. \citet{1983ApJ...266..309M}). We found the best agreement between the simulated and observed molecular column densities in the reference model, described above.

\begin{figure*}
	\includegraphics[width=1\linewidth]{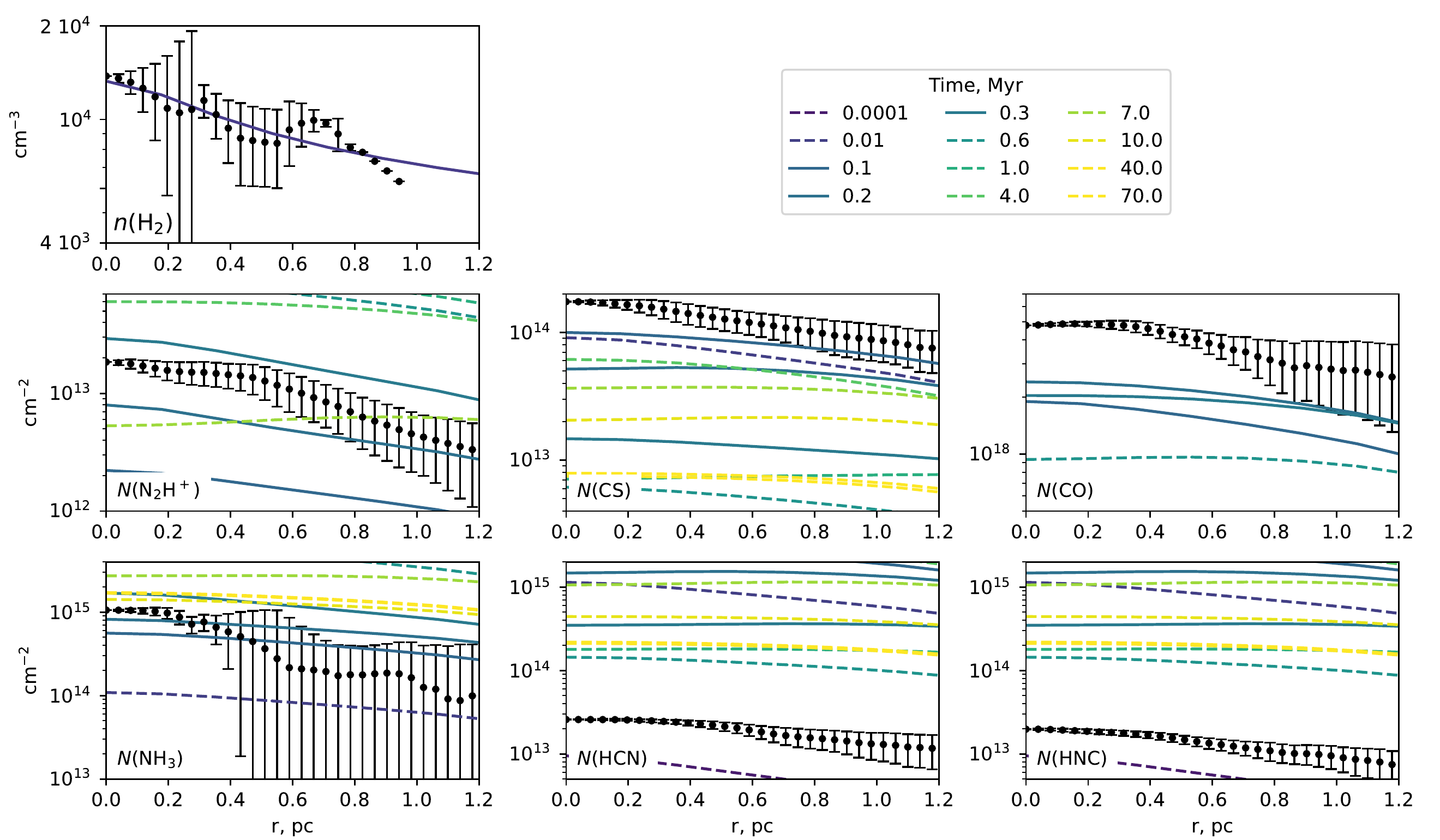} 
	\caption{Radial distributions of observed hydrogen number density $n(\rm H_2)$ and column densities of N$_2$H$^+$, CS, CO, NH$_3$, HCN and HNC in WB~673 (black lines with error bars). The panel with the $n(\rm H_2)$ graph also shows the fitted Plummer function with a dark blue line. Coloured blue-green-yellow lines on the panels with molecular column densities show results of the Presta model at different timesteps.}
	\label{fig_wb673_model}
\end{figure*}

The obtained $\Sigma$ values (see the end of Setc. \ref{chapt:mod})         from our comparison between the model and the 
         observations are shown for each clump in Fig.~\ref{fig_wb673_model}. There are several repeating patterns in the results of the simulations for all clumps in the filament. First, the computed column densities of CO and CS are lower than the observed ones over almost the entire modelled  time interval. The differences exceed an order of magnitude for clump ages exceeding several Myr for both molecules. This seemingly excludes a large chemical age for the filament. Second, simulated CS column densities agree with the observed ones to within an order of magnitude for a model time $t \leq 10^5$~yrs. Finally, simulated column densities of the N-bearing molecules NH$_3$ and N$_2$H$^+$ approach the observed values at $t\approx 2\times10^5$~yrs and at $t >10^6$~yrs. Simulated column densities of HCN and HNC molecules exceed the observed ones by $1-2.5$ orders of magnitude throughout the entire model time for WB~673. Unfortunately, the lack of data on these molecules in the other clumps prevents us from analysing them.

\begin{figure}
	\includegraphics[width=1\linewidth]{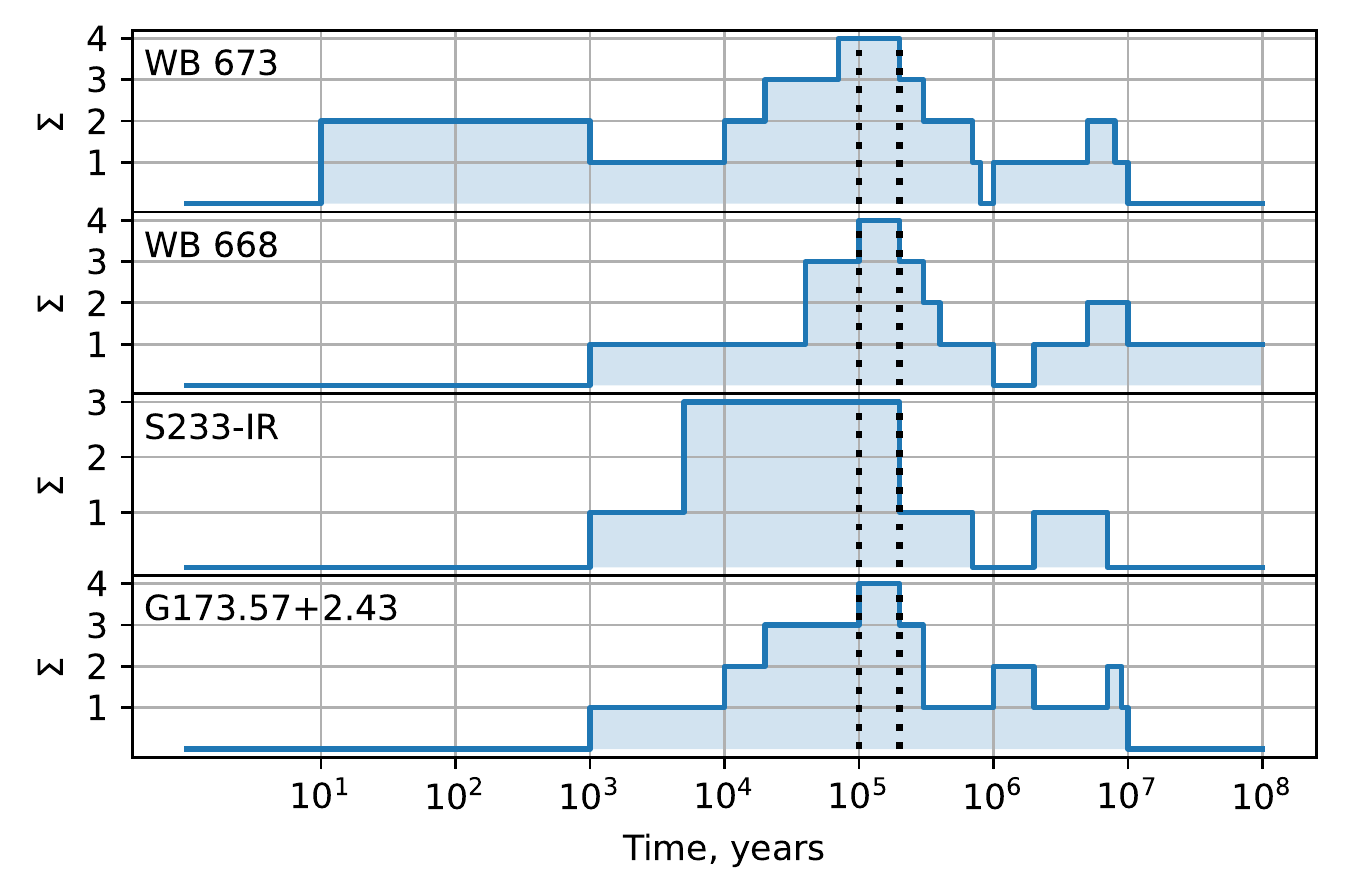} 
	\caption{The criteria of the correspondence $\Sigma$ (see the end of Sect. \ref{chapt:mod}).}
	\label{model_crit}
\end{figure}

The correspondence criteria $\Sigma$ for each clump are shown in Fig.~\ref{model_crit}. The Y-axis shows the sum of the $\Sigma$ values for all the considered molecules in the clump. $\Sigma=1$ means an agreement for just one molecule at a specific model time. $\Sigma=4$ means a full agreement for WB~668 and G173.57+2.43, but only agreement in two out of 
         three cases for WB 673, where we have data from more
         moecular species. While the periods of the best agreement and the highest $\Sigma$ are not exactly the same for the clumps, they all show the best agreement at $t = 1-3\times10^5$~yrs. Therefore, considering the clumps as parts of one single filament which was formed at some particular moment of time, we may assume this value as its chemical age. To summarise, the filament was formed rather quickly without a long previous stage of initial gas          accumulation.

\section{Discussion}

Two-phase (gas and dust) astrochemical models of molecular clouds show two regimes during the evolution of an interstellar molecular cloud, which are characterised by typical abundant species: `early' and `late' \citep[e.~g.][]{1997ApJ...486..316B, 2001ApJ...552..639A,2003ApJ...588..894S}. The models show how molecules such as CO and CS are formed rapidly in the gas phase and they become tracers of `early' chemistry. `Late' N$_2$H$^+$ and NH$_3$ molecules are formed after accretion of CO on cold dust grains. Therefore, the stage of decreasing `early' and increasing `late' molecules, which takes place in our models at the time $t=1-3\times10^5$~yrs, represents the transition between the two chemical regimes, see Fig.~\ref{fig_wb673_model}. 

Looking at the maps of $X_{\rm NH_3}$ in Fig.~\ref{fig_N}, we note higher ammonia abundances at the edges of the clumps rather than in their centres. The same is true for CS, CO, and other molecules, observed in these clumps by \citet{2020ARep...64..394R}. Depleted molecular abundances at the centre of the clumps can be attributed to their freeze-out on the surfaces of dust grains and are commonly observed in dark starless clouds \citep[e.~g.~][and many others]{2002ApJ...569..815T, 2016ApJ...830L...6J, 2022AJ....163..294P}. However, the dense clumps in this study demonstrate signs of high-mass star formation according to \citet{2016AstBu..71..208L}:  water masers at 22~GHz, molecular outflows, IRAS and MSX sources with colors corresponding to embedded \Hii{} regions in WB~673 and S233-IR and young stellar clusters at least in S233-IR~\citep[see also][]{2000A&A...361..660P}. \citet{2016AstBu..71..208L} also found proofs of gravitational instability in the clumps as virial parameters ($\alpha = M_{\rm vir}/M$)  do not exceed the critical value $\alpha_{\rm crit} = 2$  \citep{2013ApJ...779..185K}.

The discrepancy between the existence of  signs of star formation and cold chemistry in the clumps can be explained by the high density of the molecular gas surrounding the young stellar objects. Therefore, the rise of temperature, related to star formation activity is not sufficient to overwhelm the cold chemistry in the clumps. We mentioned above that the clumps are in an early stage of the star formation process. Another possible reason can be related to a deficit of Bolocam~1.1~mm emission, used to estimate $N{\rm (H_2)}$ in the present study, at the periphery of the clumps, Bolocam        measurements are not sensitive to weak widespread emission. This deficit decreases the denominator for the values of molecular relative abundances. In spite of the ambiguity in the spatial distribution of the abundances, our main results on the chemical age, rapid star formation and cold chemistry remain unchanged.

 Our conclusion on a rapid star formation time scale of $\sim 10^5$~years is consistent with several other works \citep[e.~g.~][]{2021A&A...647A.172W, 2021A&A...654A..34B}, where low- and high-mass star-forming regions were studied. On the contrary, a longer time scale $\sim 10^6$~years was found by e.~g.~\citet{2014Natur.516..219B}. Thus, chemical clocks, using the ortho-to-para ratio \citep[OPR, ][]{2014Natur.516..219B} and using abundances of `early' and `late' molecules \citep[e.~g.][and this work]{2021A&A...646A.170G, 2021A&A...647A.172W, 2021A&A...654A..34B} provide controversial results. The first type of the clock suffers from poorly known `initial' ortho-to-para ratios because this value can be $\ll 0.1$ but not 3:1 already at moderate densities $\sim 1000~\rm cm^{-3}$ at early stages of molecular cloud evolution \citep[][]{2021A&A...654L...6L}. The second type of clock suffers from unknown initial abundances of molecules in chemical models. Strictly speaking, both chemical clocks may be consistent with each other due to a relative shift of their respective zero moments. Our age may be more relevant for the `dense phase', while OPR-based ages extend to some earlier and more diffuse stages. Finally, both types of clocks suffer from poorly known and complex geometries of the objects, both in the past and at present. Therefore, each new result makes an important contribution to the statistics of the time scales.

Our result supports the idea by \citet{2015A&A...580A..49I} about formation of molecular filaments after multiple compression by supernova remnants or expanding wind-blown bubbles or \Hii{} regions. \cite{2012AJ....143...75K} found young supernova remnants whose age is $\approx 3.3\times 10^5$~yrs in the direction of G174+2.5. \citet{2017OAst...26...99K} report that the filament is situated on the border of a large and faint infrared envelope whose origin is still unknown. Therefore, we can not exclude external influence on the formation of the filament because its chemical age is similar to the age of the remnant \cite{2012AJ....143...75K}. The \Hii{} region S231 is situated on the eastern side of the filament, see Fig.~\ref{fig_wise}. We note a tighter location of star-forming regions in those parts of the filament which border the \Hii{} region. This fact indirectly supports the influence of S231 on star formation in the filament.

\section{Conclusions}

We observe and analyse ammonia emission lines toward the interstellar filament WB~673 and perform astrochemical modelling using results of the analysis in the present study. While we map the whole filament, we detect emission in the (1,1), (2,2) and (3,3) ammonia lines only in four dense clumps: WB~668, WB~673, S233-IR and G173.57+2.43. Peaks of the emission are found in the direction of the dust emission peaks at 1.1~mm in the clumps. The ammonia lines are moderately optically thick with $0.8 \leq \tau_{\rm (1,1)} \leq 1.8$ at the emission peaks. 

Using an LTE approach, we determine gas kinetic temperature, number density and ammonia column density in the dense clumps. The temperature reaches up to 30~K in the clumps. Therefore they are still cold in spite of embedded high-mass star-forming regions. The peaks of the ammonia column density in the densest parts of the clumps are almost the same, $\approx 1 - 2 \times 10^{15}$~cm$^{-2}$, based on an angular resolution of $40''$ and a  corresponding linear resolution of 0.3~pc.

Considering anomalies of the ammonia hyperfine lines, we find signatures of a collapse in WB~673. Anomalies in S233-IR correspond to a model of a medium consisting of unresolved dense clumps. We also found signatures of expansion (outflow) in S233-IR.

We reconstructed 1D density and temperature distributions in the clumps and performed their astrochemical modelling to find a chemical age of the filament. Considering four molecules: CO, CS, NH$_3$, N$_2$H$^+$ (+ HCN and HNC for WB~673), we find that the best agreement between the simulated and observed column densities reaches at $t_{\rm chem}=1-3\times 10^5$~yrs simultaneously for all the clumps. Playing with initial conditions of the chemical model, we conclude that the $t_{\rm chem}$ value represents the chemical age of the filament itself because for every clump the agreement is the best at this moment. Long preceding low-density stage of gas accumulation in the astrochemical model, breaks the agreement between the simulated and observed column densities.  Therefore, our results agree with a scenario of rapid star formation over a timescale of $\sim 10^5$~yrs. 

Internal heating sources, related to embedded high-mass young stellar objects, do not impact much on the chemistry of the dense clumps, presumably due to a high density of the surrounding material.

\section*{Acknowledgements}

We are grateful to Benjamin Winkel for calibrating the observational and to Marion Wienen for her help with the observations. We are also thankful to S.~A. Khaibrakhmanov for fruitful discussions and unknown referee for useful remarks.

The study was funded by RFBR according to research project number 20-32-90102.

\section*{Data Availability}

The data underlying this article are availabile in Zenodo at \url{https://doi.org/10.5281/zenodo.7142880}.



\bibliographystyle{mnras}
\bibliography{biblio}

%%%%%%%%%%%%%%%%% APPENDICES %%%%%%%%%%%%%%%%%%%%%

\section*{Appendix}\label{app1}

In the following we discuss the methods used for calculating rotational temperature in an optically thin case.

From the Boltzmann equation, we can obtain the relation between the temperature and the populations in the levels (J, K) and (J’, K’) \citep{1992ApJ...388..467M}:
\begin{equation}\label{Bolth}
\frac{n(J', K')}{n(J, K)} = \frac{g(J', K')}{g(J, K)}  \exp(-\frac{\Delta E (J',K'; J,K)}{T_{\rm rot}}),
\end{equation}
where g(J, K) = 2J + 1.  In the case of a homogeneous molecular cloud, the level populations are related as the column dencity of these levels:
\begin{equation}\label{ratpop}
\frac{n(J', K')}{n(J, K)} = \frac{N(J', K')}{N(J, K)}.
\end{equation}
For ammonia, the column density in an optically thin case is:
\begin{equation}\label{colden}
N(J, K) = \frac{3 h J (J + 1)}{8 \pi^3 \mu^2 k^2} \frac{1 + \exp(- h \nu / k T_{\rm ex})}{1 -  \exp(- h \nu / k T_{\rm ex})} \frac{ \int T_{\mathrm MB} dV}{(J(T_{\rm ex}) - J(T_{\rm bg}))}.
\end{equation}
Combining equations \ref{ratpop}, \ref{colden} for case (J,K) = (1,1); (J',K') = (2,2), assuming LTE ($T_{\rm ex}$ is equal): 
\begin{equation}
\frac{n(2,2)}{n(1,1)} = 3 \frac{\int {\textit T}_{\mathrm MB(2,2)}dV}{\int {\textit T}_{\mathrm MB(1,1)}dV}
\end{equation}

From equation \ref{Bolth}, the rotational temperature in the optically thin case is defined as:
\begin{equation}
T_{\rm rot} = -41.5/\ln\left(0.2\frac{\int {\textit T}_{\mathrm MB(2,2)}dV}{\int {\textit T}_{\mathrm MB(1,1)}dV}\right) (\mathrm K)
\end{equation}

% Don't change these lines
\bsp	% typesetting comment
\label{lastpage}
\end{document}